\documentclass[conference,compsoc]{IEEEtran}

\usepackage{CJKutf8}

\usepackage{amsmath,amssymb,amsfonts}
\usepackage{algorithmic}
\usepackage{graphicx}
\usepackage{textcomp}
\usepackage{xcolor}
\usepackage{booktabs}
\usepackage{csquotes}
\usepackage{float}
\usepackage{hyperref}

\begin{document}

\title{Before and after China's new Data Laws: Privacy in Apps}

\author{Anonymous}

\author{\IEEEauthorblockN{Konrad Kollnig}
\IEEEauthorblockA{\textit{Department of Computer Science}\\
\textit{University of Oxford}
}
\and
\IEEEauthorblockN{Lu Zhang}
\IEEEauthorblockA{\textit{School of Law}\\
\textit{Tsinghua University}
}
\and
\IEEEauthorblockN{Jun Zhao}
\IEEEauthorblockA{\textit{Department of Computer Science}\\
\textit{University of Oxford}
}
\and
\IEEEauthorblockN{Nigel Shadbolt}
\IEEEauthorblockA{\textit{Department of Computer Science}\\
\textit{University of Oxford}
}
}

\maketitle

\begin{abstract}
Privacy in apps is a topic of widespread interest because many apps collect and share large amounts of highly sensitive information. In response, China introduced a range of new data protection laws over recent years, notably the Personal Information Protection Law (PIPL) in 2021.
So far, there exists limited research on the impacts of these new laws on apps’ privacy practices. To address this gap, this paper analyses data collection in pairs of 634 Chinese iOS apps, one version from early 2020 and one from late 2021.
Our work finds that many more apps now implement consent. Yet, those end-users that decline consent will often be forced to exit the app.
Fewer apps now collect data without consent but many still integrate tracking libraries.
We see our findings as characteristic of a first iteration at Chinese data regulation with room for improvement.
\end{abstract}


\section{Introduction}

\noindent Tracking, the large-scale collection of data about user behaviour, is ubiquitous across mobile apps.
It is often used to make many apps available for free by showing users personalised advertising or selling their data to third parties~\cite{anirudhchi2021,mhaidli_we_2019}, and
can have disproportionate, negative effects on individuals~\cite{van_kleek_better_2017,shklovski_leakiness_2014}.
To improve the balance between protecting individuals' data and leveraging personal information in businesses and other organisations,
China has introduced various laws and regulations over recent years which govern the processing of personal data.
This includes the \emph{Personal Information Protection Law} (PIPL) from August 2021, which is the first comprehensive Chinese law on data protection~\cite{pipl_notes}.
In addition, China introduced its Data Security Law in 2021 and its first Civil Code in 2020, which put forward further regulation of data flows.
Due to the novelty of these laws, few studies have assessed their material impacts on apps' data practices.
The enactment of new data laws in China is part of wider efforts in the country to regulate the digital space.
As such, this is similar to legal initiatives in the EU, US, and UK, all of which are trying to rein in
on monopolistic behaviour of tech companies and harmful impacts of digital technologies on society.

China has a unique app ecosystem compared to other countries.
It is the only major economy in which the Google Play Store is not available, since Google does not operate many of its services in mainland China.
Regardless of that, Android (which is mainly developed by Google) has a market share of about 70\%, while iOS has about 29\%~\cite{os_china}.
On Android, there exist a range of different app stores, including those by Tencent, Oppo, Huawei, Qihoo 360, Xiaomi, and Baidu.
On iOS, the Apple App Store is the only app store.
While there exists some limited previous research on privacy in Android in China~\cite{china_2018}, no similar studies exist for iOS, despite the Apple App Store being one of the largest app ecosystems in China.
The study of iOS is especially interesting because Apple has ever more increased its market share in China over recent years~\cite{os_china}.

Motivated by the breadth of recent changes to Chinese data regulation, along with the relative absence of recent investigations into Chinese app privacy (the only previous large-scale study was done on Android and was published about five years ago, in 2018~\cite{china_2018}), this paper aims to study app privacy might have changed (particularly the implementation of consent flows) following the implementation of data regulations like the PIPL in China.
For our analysis, we draw on a body of 634 Chinese iOS apps, one version from early 2020 and one from late 2021 (i.e. after the introduction of the PIPL).
Crucially, our present analysis is not sufficient to establish whether the new laws are causally responsible for any changes in apps' privacy practices.
However, if they have indeed tackled excesses of personal data processing, we should expect at least some changes in apps' privacy practices.
Even where we do not observe any changes, it is important to characterise the status quo due to lack of previous similar studies on iOS in a Chinese context.

\section{Regulation of Data in China}

\noindent We first briefly review the history of Chinese data law, which has seen significant evolution over the past decade.

The Decision of the National People’s Congress on Strengthening the Protection of Online Information issued in 2012 is widely regarded as the starting point for Chinese data law. This has subsequently motivated Articles 1034–1039 of the China Civil Code (2020), setting forth basic rules for the protection of personal information in mainland China. The rules around data were further clarified with the Personal Information Protection Law (PIPL) from August 2021, which is the first comprehensive law to regulate the protection of personal information in mainland China. The PIPL is complemented by the Cybersecurity Law (2017) and the Data Security Law (2021), which also regulate the governance of data in the digital era. As of 9 April 2022, there were a total of 31 laws and regulations with ‘personal information’ in the title on the Peking Law System  with the keyword ‘personal information’.

Like in other countries and regions, the main purpose of Chinese data law is to balance the protection of personal information rights and the promotion of the use of personal information. The PIPL has a similar chapter structure and regulation content as the GDPR in Europe, including general provisions, rules for personal information processing, rules for cross-border personal information processing, the rights of individuals, the obligations of personal information processors, and further legal responsibilities.
Chinese data law provides seven potential legal grounds for personal information processing (PIPL Article 13). The most common legal ground for data processing, in the context of mobile apps, is ‘informed consent’. For consent to be valid, it must be voluntary, clear and fully informed (PIPL Article 14); depending on the context, further consent rules might apply.
Interestingly, there is no ‘legitimate interest’ legal ground in the PIPL, which allows data collection without consent under certain conditions under the GDPR in Europe.
As a result, much more emphasis is placed on consent in China than in Europe.

Two important further pieces of regulation are the Information Security Technology — Personal Information Security Specification (2020) (GB/T 35273-2020) (IST) and the Information Security Technology—Basic Requirements for Collecting Personal Information in Mobile Internet Applications (2022)(GB/T 41391-2022) (IST APP). These are national standards to provide detailed guidelines on personal information protection in China.
The IST provides targeted rules for various software products, while the IST APP focuses on apps only. Article 5.3 IST provides that consent should be given freely, and not against a data subject’s independent will.

The new personal data protection regulation pays much attention to the distinction between necessary and non-essential personal information, and between basic business functions and extended business functions. Article 5.3a) IST clarifies that the bundling of consent for different business functions, that require data processing, is not permitted; consent must be given to one business function at a time instead. Article 6.4.1d) IST A and Article 5.3e) IST both hold that refusing consent to one business function must not affect the use of other business functions. The specification even provides a few sample implementations of consent, including mock-ups. As such, these requirements are similar to those under Article 5(3) of the amended EU ePrivacy Directive from 2009.

Motivated by the breadth of recent changes to data regulation in China, the rest of this paper will analyse how app tracking has changed in a Chinese context since the introduction of PIPL and other notable data laws since 2020. Given the focus on consent in PIPL, we will explicitly analyse the provision and nature of consent in apps.

\section{Previous Work}

\noindent Previous research studied privacy in mobile apps extensively. Two main methods emerged: dynamic and static analysis.
\emph{Dynamic analysis} observes the run-time behaviour of an app, to gather evidence of sensitive data leaving the device.
Early research focused on OS instrumentation, i.e. modifying Android~\cite{enck_taintdroid_2010} or iOS~\cite{agarwal_protectmyprivacy_2013}.
With the growing complexity of mobile operating systems, recent work has shifted to analysing network traffic~\cite{privacyguard_vpn_2015,nomoads_2018,free_v_paid_2019,reyes_wont_2018,van_kleek_better_2017,ren_recon_2016,nomoads_2018,shuba_nomoats_2020}.
\emph{Static analysis} dissects apps without execution. Usually, apps are decompiled, and the obtained program code is analysed~\cite{han_comparing_2013,pios_2011}.
The key benefit of static analysis is that it can analyse apps quickly, allowing it to scale to millions of apps~\cite{china_2018,playdrone_2014,binns_third_2018,chen_following_2016}.
In a Chinese context, Wang et al. previously used static analysis to analyse 6 million Android apps from 16 Chinese app stores in 2018~\cite{china_2018}.
The main focus of this study was to characterise these different ecosystems and understand app security in those app stores.
These authors did not consider the Apple App Store, which is among those Chinese app stores with the largest market share~\cite{ios_marketshare_china}.
Given the increased interest in privacy by the public, regulators and lawmakers, an increasing body of literature is investigating regulatory questions~\cite{binns_measuring_2018,reyes_wont_2018,okoyomon_ridiculousness_2019,kollnig_2021,kollnig_before_2021,maps_2019,kollnig_goodbye_2022}, but hardly any work has yet covered China.

\section{Methodology}

\noindent\textbf{App download.}
For the selection of apps, we revisited the same dataset of 285,680 iOS apps as in our previous work on comparing Android and iOS privacy~\cite{kollnig2021iphones}.
These apps were selected by first generating a large list of apps available on the UK Apple App Store between December 2019 and February 2020.
Due to the global reach of the Apple App Store (including China), we noticed that the original app dataset contained a sizeable number of Chinese apps.
We selected a subset of apps that were available on the Chinese App Store.
Furthermore, we only included apps from the larger app dataset that: 1) contained Chinese (but not Japanese or Korean) characters in both their App Store title and description (which made up 7.2\% of all 285k apps) and 2) had a bundle identifier starting with \enquote{cn.} (which made up 0.4\% of all 285k apps); we made this choice to focus on apps that were developed by Chinese developers for the Chinese market.
By re-downloading those apps in October 2021, we then obtained a dataset of 634 \textit{pairs} of apps, one from before the new Chinese privacy laws and one from after.
We only included those apps that were still available on the Apple App Store in both 2020 and 2021.
We intentionally did not exclude apps that had not been updated because all apps need to comply with the new rules.

\noindent\textbf{App analysis.}
For the analysis of apps, we applied the same tools as in our previous work~\cite{kollnig2021iphones}; these tools are available online at \url{https://platformcontrol.org/}.
From this analysis, we obtained the tracking libraries integrated within apps, the tracking domains contacted upon the first app start, and the dominant companies behind this tracking and their jurisdictions; this work combines static and dynamic analysis while avoiding legal problems related to analysing iOS apps.
In contrast to our previous work, we did not study apps' sharing of PII, since Apple puts tight limits on apps' access to identifiers since the introduction of the App Tracking Transparency framework with iOS 14.5 in April 2021.

Since the new Chinese data protection law put much emphasis on consent, we additionally analysed apps' consent flows.
We additionally ran each app on a real iPhone to analyse them for consent popups.
30 seconds after having installed each app, we took a screenshot for further analysis and uninstalled it.
We inspected the screenshots for any form of 
consent, following the methodology of previous research~\cite{kollnig_2021}.
Specifically, we classified any \emph{affirmative} user agreement to data practices as consent.
While this definition of consent is arguably less strict than what is usually required under many data protection and privacy laws, this was a deliberate choice to increase the objectivity of the classification, and provide an upper bound on compliance with Chinese consent requirements.
For those apps that showed an onboarding screen at the first app start (i.e. walking users through the essentials of the app), we manually re-ran the app, tried to skip onboarding to reach the main screen, and took another screenshot for consent analysis.
We used a Chinese IP address during this analysis.

\section{Results}

\noindent In this section, we present our findings from analysing two versions~--~one from 2020 and one from 2021~--~of 634 Chinese iOS apps.

\begin{figure}
	\centering
	\includegraphics[width=\linewidth]{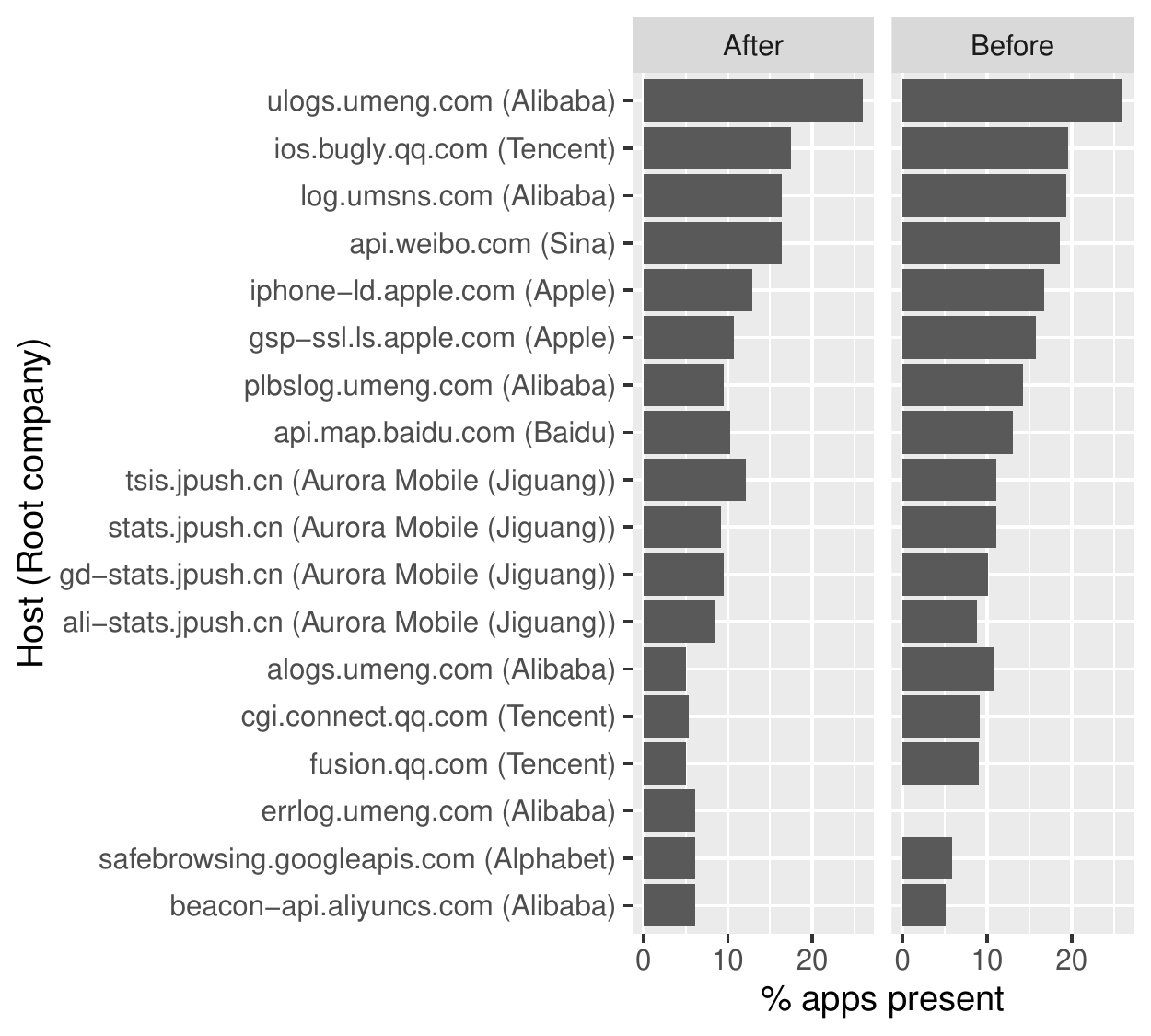}
    \footnotesize
    \begin{tabular}{lrrrrrr} \toprule
		& Median
		& Mean
		& Q1
		& Q3
		& Count $>10$
		& None \\
		\midrule
		Before & 3 & 3.7 & 1 & 6 & 4.73\% & 21.14\% \\
		After  & 2 & 3.4 & 0 & 6 & 3.94\% & 25.71\% \\ \bottomrule
	\end{tabular}
	\caption{Top 15 tracking hosts contacted at the first app start and without consent, as well as the companies owning them.}
	\label{fig:tracker_hosts}
\end{figure}

\subsection{Contacted Trackers without Consent}
\label{sec:data_sharing_consent}

\noindent This section analyses how many tracking domains are contacted by the studied apps before any user interaction takes place.
Since tracking libraries usually start sending data right at the first app start~\cite{kollnig_2021,reyes_wont_2018,nguyen_share_first_consent_2021,kollnig2021iphones}, this approach provides additional evidence as to the nature of tracking in apps~--~and without consent.
Our results are shown in Figure~\ref{fig:tracker_hosts}.

\begin{figure}
    \centering
    \includegraphics[width=0.7\linewidth]{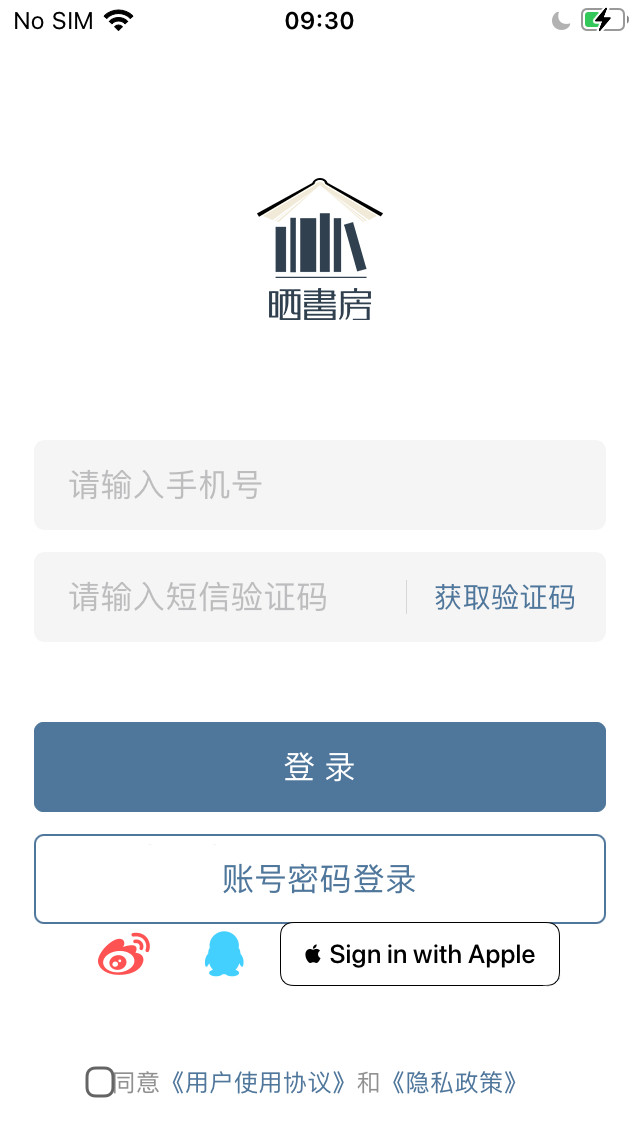}
    \caption{A typical login screen for a Chinese iOS app. Login is possible through a range of messaging and social media services. For example, the penguin belongs to Tencent's QQ instant messaging software service.
    The red icon belongs to Weibo, a social media service, which was the second most commonly contacted service upon the first app initiation (see Section~\ref{sec:data_sharing_consent}).
    Some apps also allow the sign-in with an Apple account.
    Before logging in, users must first check the checkbox and thereby agree to the Terms and Conditions and the Privacy Policy (i.e. an `unticked checkbox' type of consent in Section~\ref{sec:consent}).}
    \label{fig:app_login}
\end{figure}

The average number of tracking domains contacted decreased somewhat (3.7 before, 3.4 after).
The number of different contacted tracking companies also declined (from 2.2 before to 1.9 after).
About one quarter of apps, from both before and after the new laws, did not contact any tracking domains at the first app start.
The most popular domain is related to Alibaba's analytics services at \texttt{ulogs.umeng.com} (25.9\% of apps before, 26.0\% after).
This is followed by Tencent's Bugly at \texttt{ios.bugly.qq.com} (19.6\% of apps before, 17.5\% after), Alibaba's \texttt{log.umsns.com} (19.4\% of apps before, 16.4\% after), and Weibo at \texttt{api.weibo.com} (18.6\% of apps before, 16.4\% after).
Weibo is one of the biggest social media platforms in China, and is often used as a login for mobile apps (see Figure~\ref{fig:app_login}).
The fact that social media and messaging services are commonly used for authentication with apps is likely responsible for some of the observed data collection without consent.

\begin{figure}
	\centering
	\includegraphics[width=\linewidth]{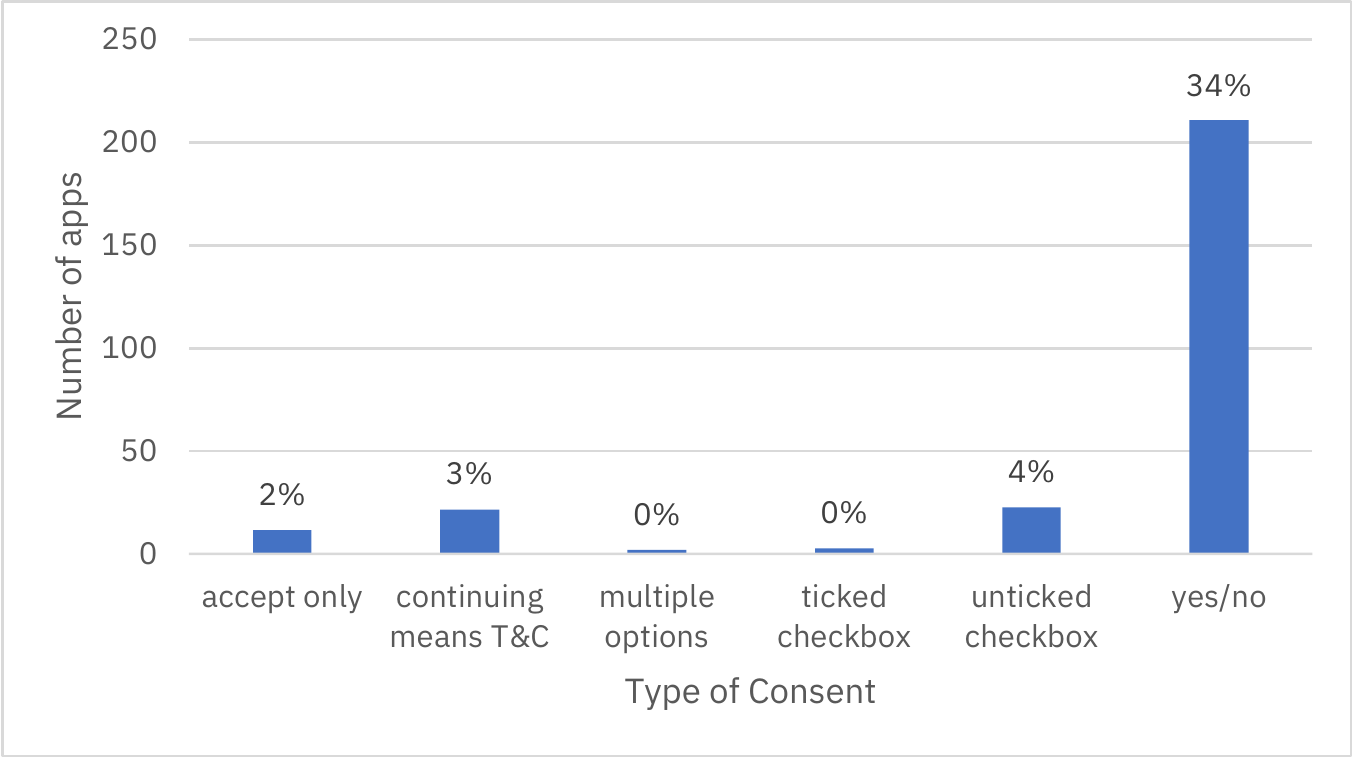}
	\caption{Types of consent in apps. In our analysis, we applied a broad definition of consent, including all types of affirmative user choice over data in apps; this is to improve the objectivity of our analysis.
	The Appendix provides a screenshot for each consent type.}
	\label{fig:consent}
\end{figure}

\subsection{Provision of Consent}
\label{sec:consent}

\noindent Some of the analysed apps had problems showing a UI (e.g. crashed or just showed a black screen), so we excluded 34 from the following consent analysis, leading to 604 remaining apps.
98 apps (16.2\%) showed an onboarding screen, so we re-ran these apps and skipped the onboarding. 169 (28.0\%) asked for login credentials (a common example is shown in Figure~\ref{fig:app_login}) and might obtain consent from users elsewhere.

A total of 274 apps from the 2021 dataset (45.4\%) asked for consent. 145 apps (24.0\% of all apps, 52.9\% of apps with consent in 2021) added consent notices compared to their 2020 version.
3 apps did not show consent in 2021 but did so in 2020.
Among apps that did not show a login screen, 215 apps (49.4\% of apps without a login screen) asked for consent; 59 apps with a login screen (34.9\% of apps with a login screen) asked for consent. The discrepancy between the percentages indicates that we miss some consent flows for apps with a login screen, but still detect a sizeable share.

We also analysed in what way apps asked users for consent, see Figure~\ref{fig:consent}. 212 apps (77.4\% of apps that asked for consent) showed a binary choice. These apps usually ask in a popup screen whether the user agrees to the privacy policy or the terms of use, and exit the app on refusal. 12 apps (4.4\%) only allowed users to accept and did not display a refusal option. 2 apps (0.7\%) showed more options than just a binary choice.
23 apps (8.4\%) showed an unticked checkbox for consent, 3 (1.1\%) a pre-ticked checkbox.
22 apps (8.0\%) indicated that continuing to use the app (e.g. logging in) would mean acceptance of the terms of use or privacy policy (\enquote{continuing means T\&C} in Figure~\ref{fig:consent}). The Appendix provides a screenshot for each consent type.

Our analysis of consent in apps points to a relative absence of granularity in consent implementations.
As such, some of these apps might conflict with Article 5.3 of the Information Security Technology — Personal Information Security Specification. This specification requires separate consent for different business functions, that such consent is freely given, and that refusing consent does not have negative effects on the use of other business functions.
In the case of apps, this could mean that apps need to provide more granular consent options and may not be allowed to exit the app if consent is refused.
The Specification distinguishes between basic business functions and extended business functions. The former refers to the basic expectation and the most important demand of users to use the products or services provided, while the latter refers to other functions other than the basic business functions provided by the products or services.
If the personal information subject does not agree to collect the personal information necessary for extended business functions, it shall not refuse to provide the basic business function or reduce the service quality of the basic business function.
However, the line between different business functions as defined by the specification depends on each specific app.
The Specification even explicitly states in Article 5.3f) that solely for reasons such as improving service quality, enhancing user experience, developing new products and enhancing security, individuals may not be compelled to consent to data collection activities.
Article 11 of the Data Security Law further points out that network operators shall not force or mislead personal information subjects to consent in the form of default authorization or function bundling under the pretext of improving service quality, enhancing user experience, pushing targeted  information or developing new products, and similar purposes.

\begin{figure}
	\centering
    \includegraphics[width=\linewidth]{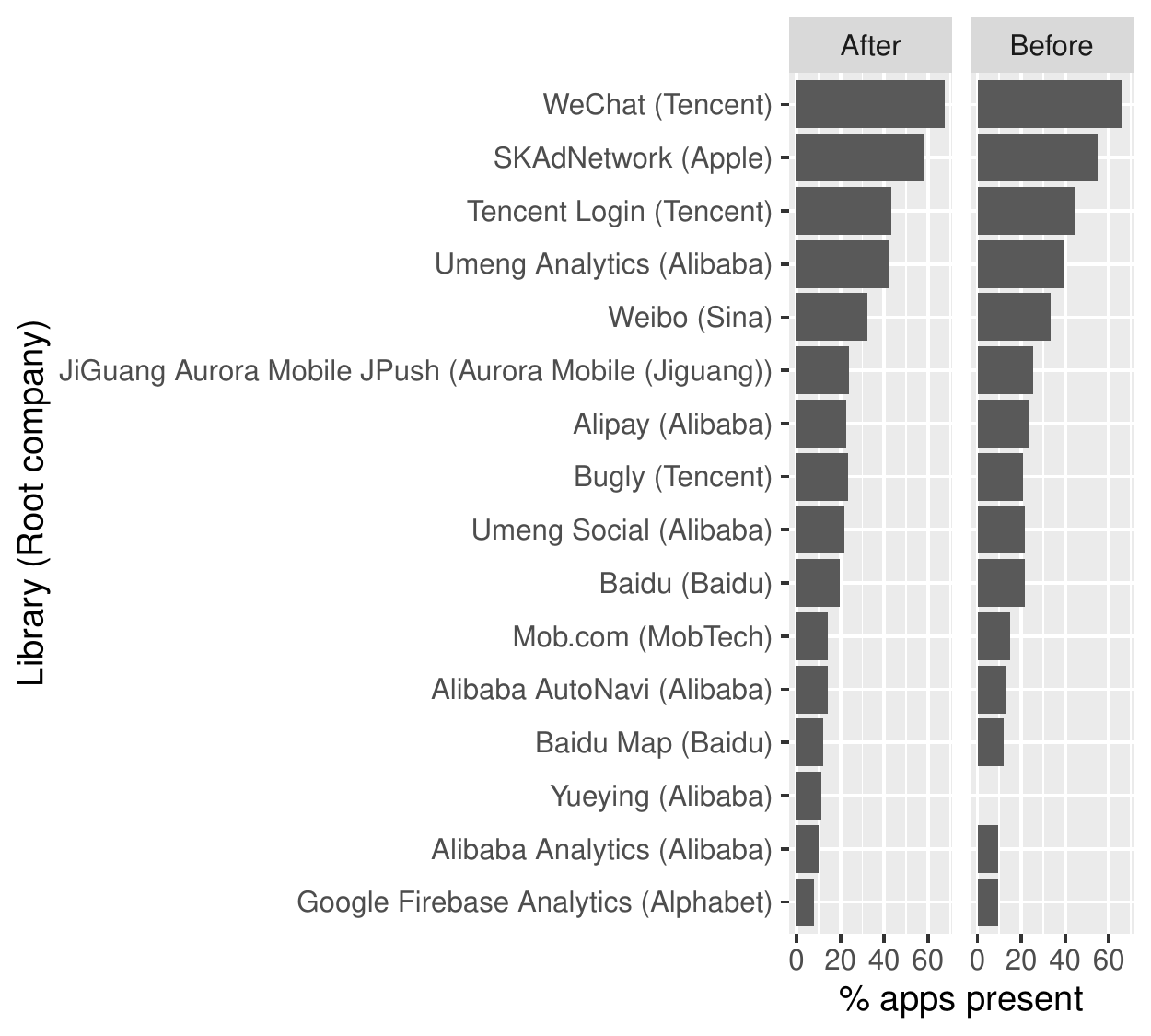}
    \centering
    \footnotesize
    \begin{tabular}{lrrrrrr} \toprule
		& Median
		& Mean
		& Q1
		& Q3
		& Count $>10$
		& None \\
		\midrule
		Before & 4 & 4.7 & 2 & 7 & 3.47\% & 8.20\% \\
		After  & 5 & 4.9 & 2 & 7 & 3.94\% & 7.89\% \\ \bottomrule
	\end{tabular}
	\caption{Top 15 third-party libraries, as well as the companies owning them (in brackets).}
	\label{fig:tracker_libraries}
\end{figure}

For comparison, in an EU context, our previous work on consent in a representative sample of 1,297 Google Play apps from 2020 found that only 9.9\% of apps asked for any form of consent~\cite{kollnig_2021}.
The observed share of Chinese apps with consent flows is much higher.
It is, however, difficult to compare the two studies because of differences between the app stores and sampling techniques.

\subsection{Tracking Libraries in Apps}
\label{sec:static_tracking}

\noindent Apps from both before the Chinese data protection laws and after widely integrated tracking libraries
(see Figure~\ref{fig:tracker_libraries}). The median number of tracking libraries included in an app was 4 before and 5 after.
The mean before was 4.7, and the mean after was 4.9.
3.47\% of apps from before contained more than 10 tracking libraries, compared to 3.94\% after. 91.80\% contained at least one before the new laws, and 92.11\% after.

The most prominent libraries have not changed since the introduction of the new laws.
The top one remains WeChat (in 65.8\% of apps before, and 67.8\% after). This is followed by Apple's SKAdNetwork library (54.9\% before, 58.0\% after). While part of Apple's privacy-preserving advertising attribution system, this library discloses information about what ads a user clicked on to Apple, from which Apple could (theoretically) build user profiles for its own advertising system~\cite{kollnig_goodbye_2022}.
Tencent Login ranks third (43.2\% before, 44.5\% after), closely followed by Umeng Analytics (39.9\% before, 42.6\%).

\begin{figure}
    \centering
    \includegraphics[width=0.7\linewidth]{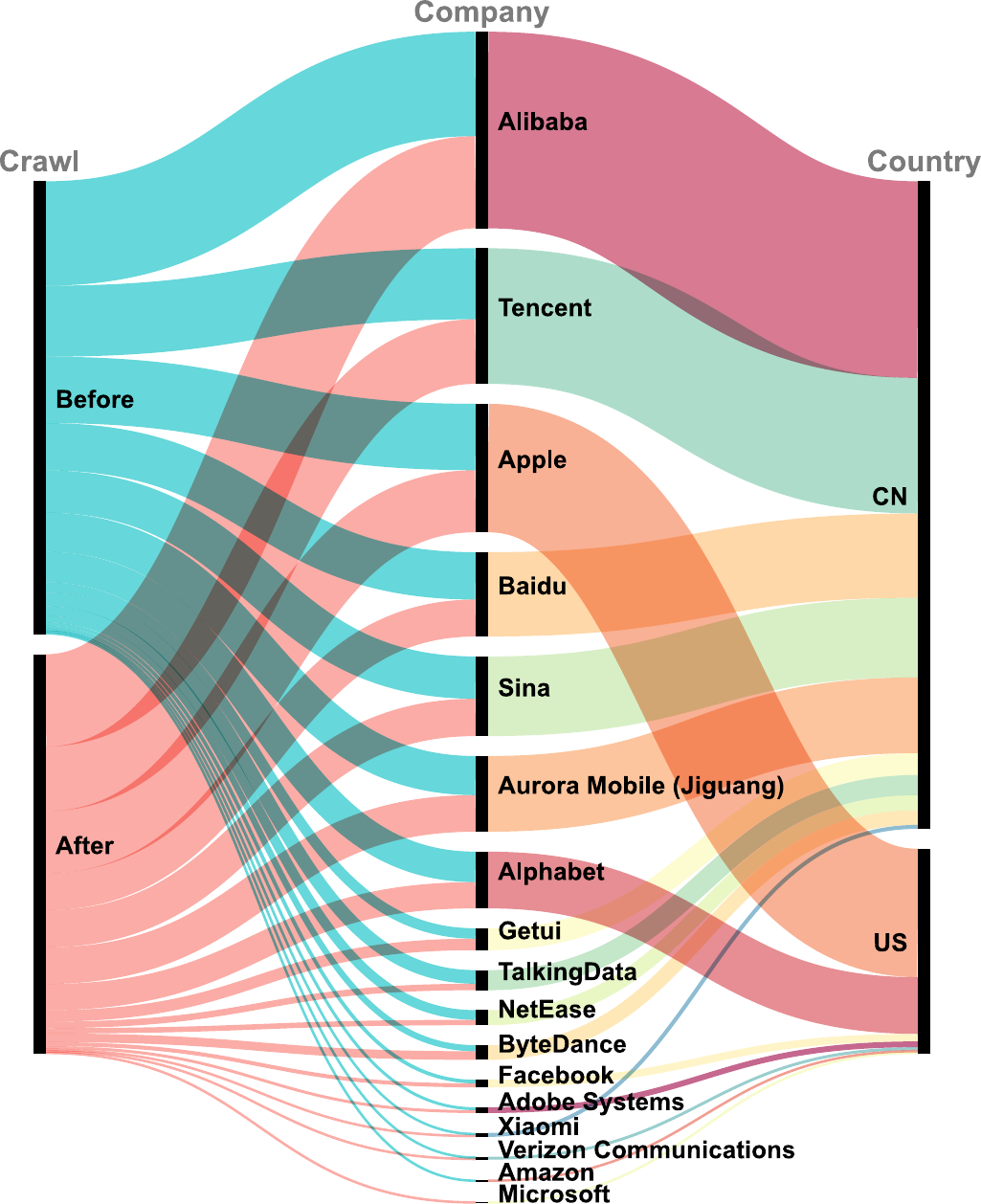}
    
     \vspace{0.2cm}
     
        \footnotesize
    \begin{tabular}{l r r r r r r} \toprule
		& Median
		& Mean
		& Q1
		& Q3
		& None \\
		\midrule
		Before & 4 & 3.6 & 2 & 5 & 5.52\% \\
		After  & 4 & 3.6 & 2 & 5 & 5.21\% \\ \bottomrule
	\end{tabular}

    \caption{Visualisation of third-party tracking  across root companies and their jurisdictions, in 2020 \& 2021.
    This information was derived both from contacted domains and included libraries.
    Most of the data collection from Chinese apps goes to Chinese companies. The only exceptions are Apple and to a much lesser extent Alphabet/Google, both based in the US. However, Google does not operate in China and does not usually collect data from mainland China (though in Hong Kong).}
    \label{fig:alluvial}
\end{figure}

\subsection{Companies behind Tracking Technology}
\label{sec:tracking_companies}

\noindent Since some tracker companies belong to a larger consortium of companies,
we now consider what parent companies ultimately own the tracking technology, i.e. the \textit{root companies} behind tracker companies.
We report these root companies by combining the observations from our static and traffic analysis.
This is visualised in Figure~\ref{fig:alluvial}.
Apple stands out as the most dominant foreign company with a share in iOS data collection in China.
The median number of companies was 4 both before and after the new laws. The mean was and remains at 3.6.
The maximum number of companies was 10 before and after the new laws.
About 75\% of apps could share data with Tencent (the parent company of WeChat and QQ), closely followed by Apple.
The next most common is Alibaba, which could be contacted by about two-thirds of the analysed apps.
Beyond the top three, a range of further companies 
collect data from apps, including Sina (a Chinese tech company and developer of the microblogging service Weibo), Baidu (the company behind the leading search engine), and Aurora Mobile (a company -- also known as Jiguang -- that offers similar services to Google Firebase, including ads, push notifications and analytics).

\section{Conclusions}
\noindent This paper analysed data collection in pairs of 634 Chinese iOS apps. These apps stemmed from before and after the introduction of key Chinese data privacy laws, particularly the PIPL from 2021. Our research aimed to track the changes over time and since the new laws.

In our analysis, we noticed a discrepancy between current legal norms in China and data practices within apps.
Analysing consent in apps, we noticed that such was often restricted to yes/no selections and that choosing not to agree to data practices commonly exits the app. Thus, many apps do not offer the level of granularity required under Chinese law.
In practice, bundling of consent was common and there was often no distinction between necessary personal information and non-essential personal information, nor a distinction between basic business functions and extended business functions.
Furthermore, it is often hard to withdraw their consent after consenting to data collection within an app. The three types of consent in law (general consent, individual consent and written consent) were also not reflected in practice.
However, we observed that many apps have now added consent flows into their existing apps at the first app start: 45.5\% of the 2021 apps asked for some form of consent in our analysis, compared to 21.8\% in 2020.

We further found that Chinese apps, that have been available on the Apple App Store since 2020 or longer, still integrate a similar number of tracking libraries.
Yet, the number of tracking companies contacted upon the first app start and without consent declined.

What we see from our results is arguably the \textit{first iteration} of data regulation in Chinese apps. It is not perfect and there remains room for improvement. This particularly concerns the quality of consent in many of the studied apps.
As regulators around the world ramp up enforcement and shape norms around apps’ data practices, apps’ practices will change and evolve.
This will happen particularly in the Chinese context with active regulatory organisations that are highly interested in mobile apps.
We should expect a second wave of app regulation over the coming years, not necessarily through new legislation, but rather through evolving norms, practices, and consumer expectations around apps.

\section*{Acknowledgment}
\noindent We thank Michael Cerny for his helpful comments.
Konrad Kollnig was funded by the UK Engineering and Physical Sciences Research Council (EPSRC) under grant number EP/R513295/1.
Konrad Kollnig, Jun Zhao and Nigel Shadbolt have been supported by the Oxford Martin School EWADA Programme.

\bibliographystyle{IEEEtran}
\bibliography{references}

\section*{Details on Tracker Library}

\begin{figure}[H]
\centering
\begin{tabular}{@{}ll@{}}
\toprule
Library                     & Main Functionality                            \\ \midrule
WeChat                      & Login, Social                              \\
SKAdNetwork                 & Advertising Attribution                       \\
Tencent Login               & Login                                         \\
Weibo                       & Login, Social                                      \\
JiGuang & Push, Analytics, Social \\
Alipay                      & Payment                                       \\
Bugly                       & Crash Reporting                 \\
Umeng Social                & Social                                \\
Baidu                       & Analytics, Ads, Maps                          \\
Mob.com SDK                     & Push, Social, Ads            \\
Alibaba AutoNavi            & Maps                                          \\
Baidu Map                   & Maps                                          \\
UC Yueying                     & Crash Reporting                                     \\
Alibaba Analytics           & Analytics                                     \\
Google Firebase Analytics   & Analytics \\
\bottomrule                       
\end{tabular}
\caption{The main functionality of the identified top 15 tracking SDKs. Many of the identified libraries are multi-purpose SDKs and offer a range of functionality.}
\end{figure}

\newpage

\section*{Observed Types of Consents}
\label{sec:consent-examples}
\noindent This section provides examples for the different types of consent we observed in analysing apps in Section~\ref{sec:consent}; this analysis is visualised in Figure~\ref{fig:consent}.

\vspace{2cm}

\begin{figure}[H]
	\centering
	\includegraphics[height=0.35\textheight]{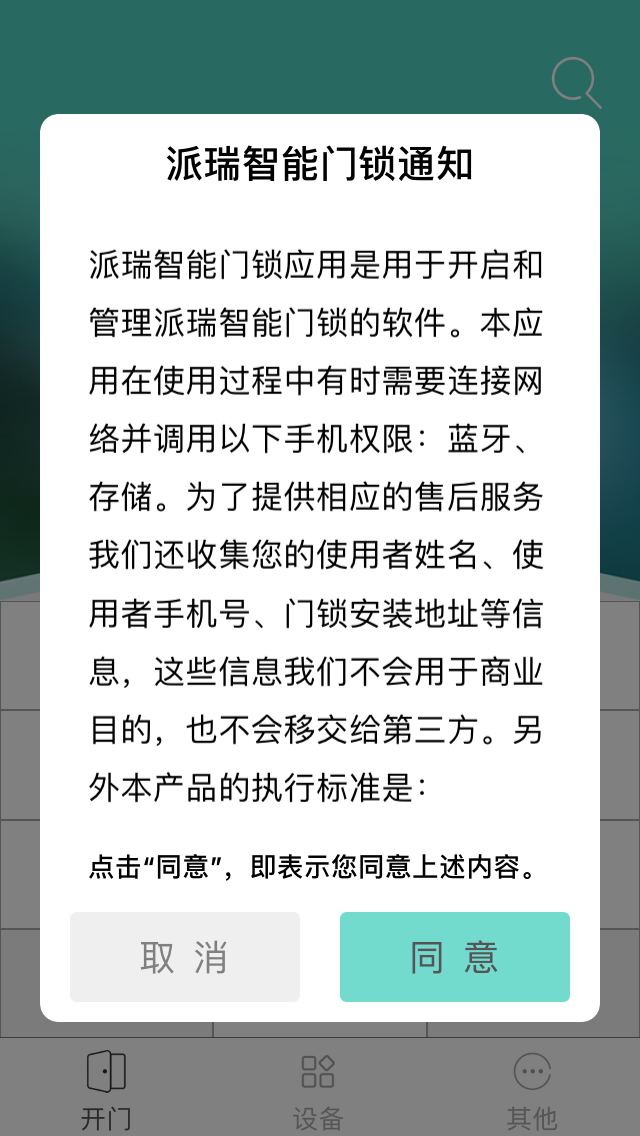}
	\caption{An example for the \enquote{yes/no} type of consent in Figure~\ref{fig:consent}.}
\end{figure}

\begin{figure}[H]
	\centering
	\includegraphics[height=0.35\textheight]{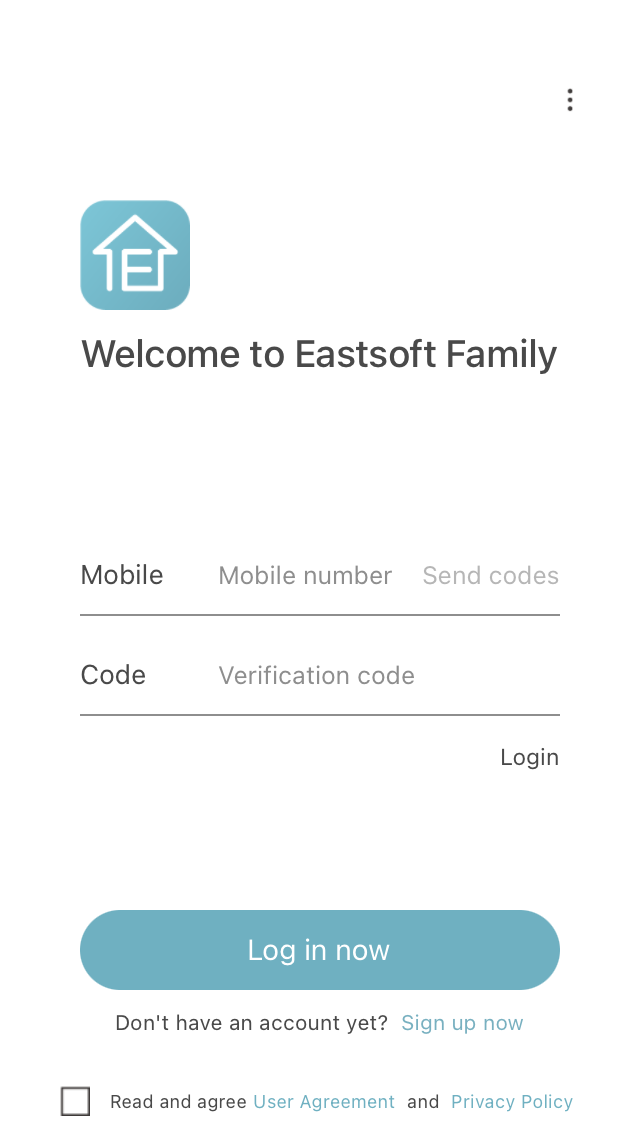}
	\caption{An example for the \enquote{unticked checkbox} type of consent in Figure~\ref{fig:consent}.}
\end{figure}

\begin{figure}[H]
	\centering
	\includegraphics[height=0.35\textheight]{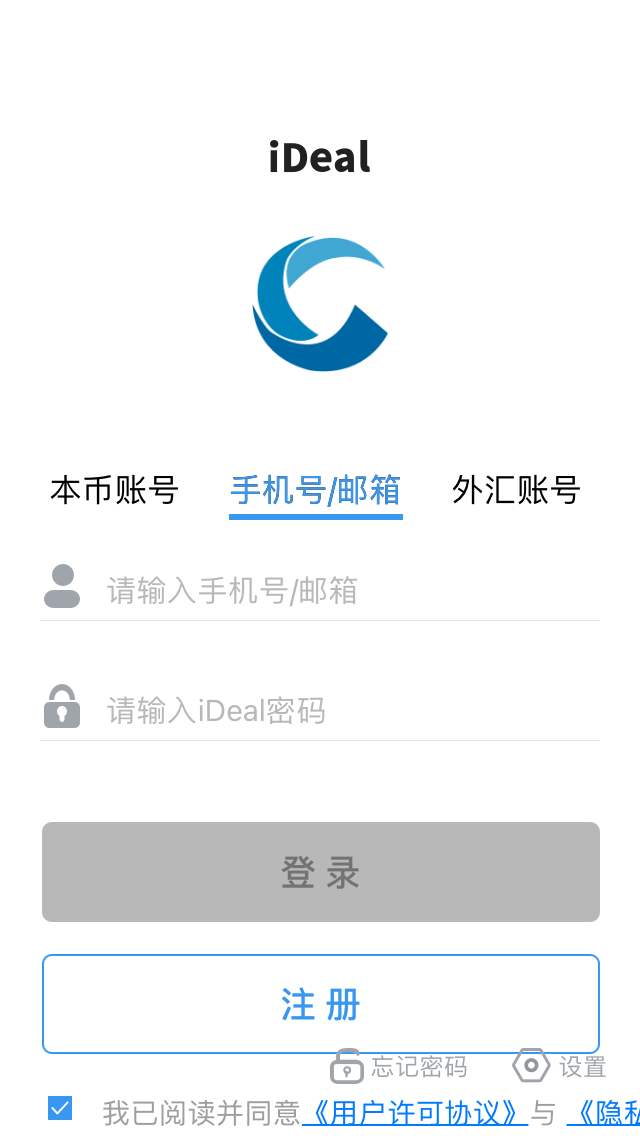}
	\caption{An example for the \enquote{ticked checkbox} type of consent in Figure~\ref{fig:consent}.}
\end{figure}

\begin{figure}[H]
	\centering
	\includegraphics[height=0.35\textheight]{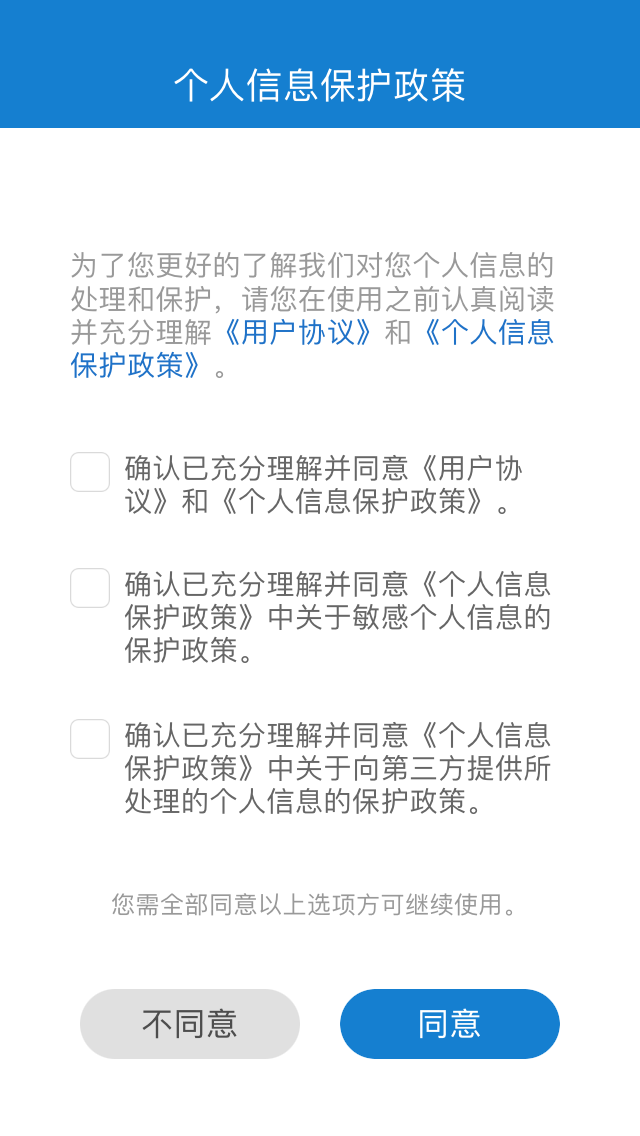}
	\caption{An example for the \enquote{multiple options checkbox} type of consent in Figure~\ref{fig:consent}.}
\end{figure}

\begin{figure}[H]
	\centering
	\includegraphics[height=0.35\textheight]{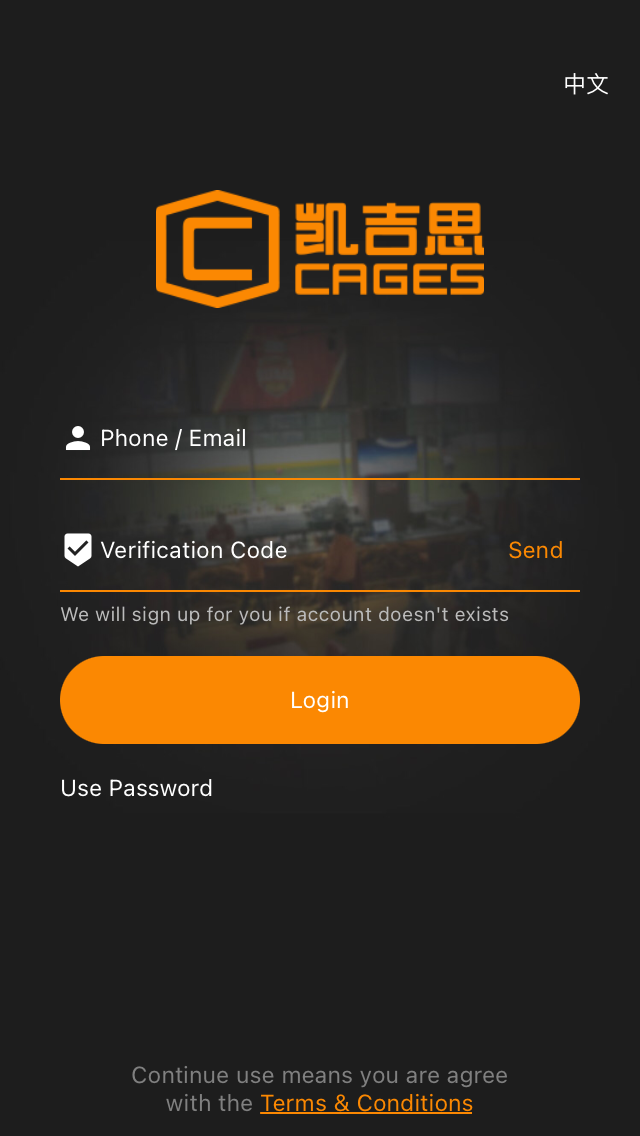}
	\caption{An example for the \enquote{continuing means T\&C} type of consent in Figure~\ref{fig:consent}.}
\end{figure}

\begin{figure}[H]
	\centering
	\includegraphics[height=0.35\textheight]{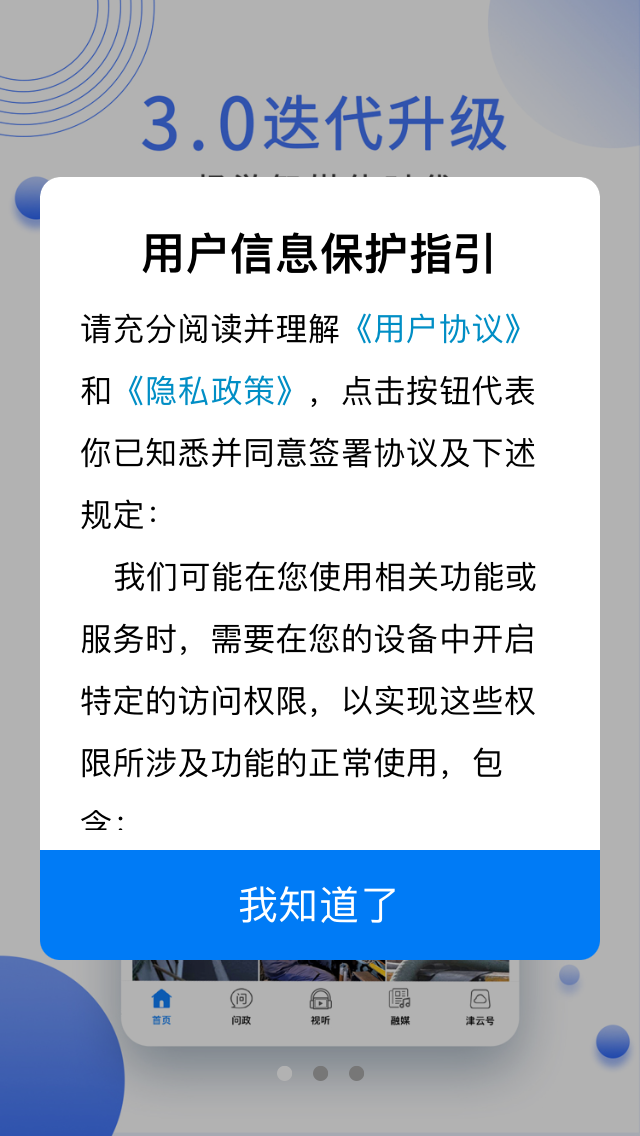}
	\caption{An example for the \enquote{accept only} type of consent in Figure~\ref{fig:consent}.}
\end{figure}

\end{document}